\documentclass[sigconf]{acmart}

\usepackage{balance}
\usepackage{subcaption}
\usepackage{microtype}

\usepackage{acmart-taps}

\AtBeginDocument{%
  }

\copyrightyear{2026}
\acmYear{2026}
\setcopyright{cc}
\setcctype{by}
\acmConference[ACE 2026]{28th Australasian Computing Education
Conference}{February 09--13, 2026}{Melbourne, VIC, Australia}
\acmBooktitle{28th Australasian Computing Education Conference (ACE
2026), February 09--13, 2026, Melbourne, VIC, Australia}
\acmPrice{}
\acmDOI{10.1145/3786228.3786237}
\acmISBN{979-8-4007-2352-0/2026/02}

\begin{document}

\title{LLMs Unplugged:\\Teaching Resources for a ChatGPT World}

\author{Ben Swift}
\email{ben.swift@anu.edu.au}
\orcid{0000-0003-2138-5969}
\affiliation{%
  \institution{Australian National University}
  \city{Canberra}
  \state{ACT}
  \country{Australia}
}

\begin{abstract}
  Large Language Models (LLMs) are everywhere, yet many learners lack a concrete
  mental model of how they generate text. This paper presents
  \emph{LLMs Unplugged}, an unplugged set of activities that teaches
  the training-to-generation
  loop (and beyond) using hand-built n-gram models and simple
  weighted sampling. Workshops
  based on these resources have been delivered to over 400 participants
  across secondary, tertiary, and executive-education contexts, and participants
  report that the activities demystify LLMs by reframing them as probabilistic
  "next word generation" at scale. All resources are freely
  available under a Creative Commons license at
  \url{www.llmsunplugged.org}, with a modular design that supports
  anything from a one hour crash course to a several-day intensive workshop.
\end{abstract}




\maketitle

\section{Introduction}

For over two decades \emph{CS Unplugged} has shown that core computing
concepts can be taught effectively without computers
\citep{bellCSUnpluggedHow2018}. Through carefully designed hands-on
activities, learners from primary school to adult education have learned
about algorithms, data structures, and computational thinking. The
approach strips away the distractions of syntax and tooling, allowing
learners to focus on underlying principles. It works
\citep{chenFosteringComputationalThinking2023}, and makes learning this
stuff \emph{fun}.

As Machine Learning (in the 2010s) and Artificial Intelligence (in the
2020s) have become more prominent in public discourse, educators have
naturally extended the unplugged approach to these CS subfields
\citep{lindnerUnpluggedActivitiesContext2019}. There are unplugged
activities for teaching classification, clustering, computer vision and
artificial neural network (ANN) concepts, showing that even
sophisticated AI ideas can be made tangible through physical activities
\citep{songArtificialIntelligenceUnplugged2024}.

However, the public release of ChatGPT in November 2022
\citep{openaiChatGPT2022} shifted what ``AI'' means to most people.
Large language models (LLMs) moved from research curiosity to ubiquitous
tool almost overnight. Within months, knowledge workers across every
domain were using LLMs daily \citep{liaoLLMsResearchTools2025}, most of
whom had no real mental model of how the text they typed into the
ChatGPT prompt box produced the text they received as a response.

The existing collections of CS and AI unplugged activities
\citep{csunplugged} \citep{seegererAIUnplugged, northwesternAIUnplugged}
do not contain many resources specifically about language models or text
generation. \citep{northwesternAIUnplugged} has one ``Large Language Mad
Libs'' activity, but it involves students actually using ChatGPT.
\citep{connellyLLMsBlackBoxes2025} has a ``cut sentences into
words/tokens'' activity, but the text generation involves drawing the
cut-up words from the bag at random---not a process conducive to
high-quality text generation.
\emph{CS In Schools} \citep{williamsCSInSchools2020} has one activity on
Generative AI that uses a ``counting bigrams'' approach to analysing
text, but doesn't show how to generate new text.

This gap is particularly acute because LLMs have become pervasive far
beyond traditional computing education contexts
\citep{zhangSystematicReviewChatGPT2024}. School-age students certainly
need to understand these tools, but so do public servants, executives,
journalists, and anyone else who interacts with the world around them
via an LLM interface (which increasingly looks like most of us, whether
we like it or not). The unplugged approach---building understanding
through hands-on activities rather than abstract explanation---seems
ideally suited to this broad audience.

This practitioner paper presents \emph{LLMs Unplugged}, a
ready-to-use unplugged teaching resource for teaching LLM fundamentals,
including:

\begin{itemize}
  \item
    a curriculum covering the complete training-to-generation pipeline
    for language models through hands-on activities
  \item
    a modular lesson structure: two Fundamentals lessons
    (\emph{Training} and \emph{Generation}) and Extensions for deeper
    exploration
  \item
    open materials (CC BY-NC-SA): an online portal at
    \url{www.llmsunplugged.org} with lessons with printable
    handouts (see Figure 1), instructor notes, and software tools
    (either web-based or CLI, MIT-licensed) to produce
    domain-specific n-gram booklets
  \item
    practitioner insights from approximately 400 participants across
    diverse audiences, summarising qualitative reception and delivery
    patterns
\end{itemize}

\section{What might ``LLMs unplugged'' look like?}

Language modelling has deep historical roots. Andrey Markov's 1913 work
on stochastic processes applied to letter sequences in Pushkin's
\emph{Eugene Onegin} \citep{linkChainsWestMarkovs2006} established the
mathematical foundation for modelling language as sequences of dependent
random variables. Markov's interest was purely mathematical---proving
properties of dependent random variables through empirical text
analysis---but his work established that language has statistical
structure that can be quantified.

Claude Shannon built directly on this foundation three decades later
\citep{shannonPredictionEntropyPrinted1951}. Between 1948 and 1951, Shannon
applied his new information theory to written English, using n-gram
models to measure entropy and redundancy in language. Crucially, Shannon
was the first to systematically generate synthetic text using these
models, starting with random letters (0-gram), then letter frequencies
(1-gram), then letter pairs (2-gram), and progressively higher orders.
This generative approach revealed how increasing context length produces
increasingly realistic text---a finding that remains central to modern
language models.

Markov and Shannon's work was itself ``unplugged'': counting transitions
by hand, calculating probabilities manually, and even generating
synthetic text by creating hand-drawn tables and selecting letters based
on their frequencies. Modern LLMs use the same fundamental
approach---modelling language as weighted distributions over
sequences---but at vastly greater scale and with learned rather than
hand-crafted statistics. This historical work was influential in the design
of \emph{LLMs Unplugged}.

When considering how to teach the fundamental concepts of language
models in an unplugged style, several design constraints emerged from
\emph{CS Unplugged} design patterns \citep{nishidaCSUnpluggedDesign2009} and
the specific characteristics of language models.

\begin{itemize}
  \item
    End-to-end generation: cover the complete
    training \textrightarrow{} generation pipeline. Language models are
    fundamentally generative; preserving this quality keeps activities
    engaging and directly relevant to how people use LLM tools.
  \item
    Modular and low-friction: each activity should stand alone in a short
    session, need minimal materials, work for various group sizes, and be
    easy to adapt. Lessons should build on each other but also be
    independently useful.
  \item
    Broad accessibility: it needs to work for audiences beyond
    computing students. No
    programming assumed and minimal mathematics; rely on hands-on
    activities and plain language.
\end{itemize}

\section{LLMs Unplugged}

The \emph{LLMs Unplugged} suite of unplugged activities is
available under a Creative Commons license (CC-BY-NC-SA) from
\url{www.llmsunplugged.org}.

The core mechanic is simple: students build their own n-gram
language models (from scratch) using a children's book such as
\emph{Dick and Jane} \citep{gray1946fun} or \emph{Dear Zoo}
\citep{campbell1982dear} as the training text. In the training phase
students fill out this grid by hand, tracking which words follow which
other words through simple tally marks. In the generation phase they use
their newly trained ``model grid'' to iteratively generate text by
looking up all possible \emph{next} words (and their relative
frequencies) and selecting one at random with a dice roll.

The primary learning outcome is straightforward:
students understand that language models---whether ChatGPT or their
hand-built version---work the same way. LLMs work by keeping track of
the patterns in existing text, then generate new text by repeatedly
making random choices weighted by what they've seen before.

The \emph{LLMs Unplugged} website linked above has three
main components:

\begin{enumerate}
  \item
    The \textbf{lessons} are the core content: each is a self-contained
    activity which establishes the context (what you'll need, your goal,
    the key idea), describes the algorithm/procedure, and gives a worked
    example. Printable handouts are also available for classroom use. See
    the lesson progression in Section~\ref{sec:lesson-progression}
    for a full list of the
    concepts covered in each lesson.
  \item
    The \textbf{instructor notes} provide the pedagogical scaffolding. For
    each lesson, they explain the connection between the activity and
    modern LLMs, suggest discussion questions to deepen understanding, and
    provide historical or technical context. These notes help educators
    without deep AI expertise deliver the material effectively.
  \item
    (Optional) The resources include an open-source (MIT Licensed)
    \textbf{software tool} to allow educators to create custom n-gram
    booklets from any text corpus. The tool is written in Rust
    \citep{matsakisRustLanguage2014} and uses Typst
    \citep{maedjeTypstProgrammableMarkup2022} for typesetting, but
    there is also a web-based version at
    \url{www.llmsunplugged.org/tools}. For any
    input text/pdf/docx file the tool will tokenise it, compute
    n-gram statistics,
    and create a formatted n-gram booklet. This means educators can
    pre-train models on domain-relevant text---medical case studies, legal
    documents, poetry, student essays; whatever best connects with their
    participants---which can then be used in (almost) all of the lesson
    activities. Using these software tools is optional; the activities
    work perfectly well when hand-trained on short texts. But they enable
    scaling the approach to longer texts and larger vocabularies without
    requiring students to spend hours tallying word pairs, while still
    allowing for ``unplugged'' text generation.
\end{enumerate}

Example lesson handouts for the \emph{Training},
\emph{Generation} and \emph{Pre-trained model generation} are
included in an Appendix at the end of this paper; all other handouts
are available from the LLMs Unplugged website.

\subsection{Lesson progression}
\label{sec:lesson-progression}

The \emph{LLMs Unplugged} lessons are organised into
Fundamentals---which should be completed in order---and
Extensions that
can be selected based on interest and available time.

\subsubsection{Fundamentals}

These two lessons build on each other and form the core of any \emph{LLMs
Unplugged} workshop.

\begin{itemize}
  \item
    \emph{Training}: students process text by hand---converting to
    lowercase, treating punctuation as tokens, then counting word pairs
    and recording tallies in a grid. This mirrors the core of LLM
    training: learning is counting patterns in text, and the resulting
    grid \emph{is} the model.
  \item
    \emph{Generation}: students use dice rolls weighted by their grid's
    counts to sample next words one at a time. The randomness explains
    why LLMs give different responses to the same prompt---and why
    the output is often predictable, but sometimes surprising.
\end{itemize}

These Fundamentals lessons are
available in two
variants: \emph{grid} and \emph{bucket}. The grid version described
above uses paper
grids and dice rolls for weighted sampling, connecting well to
probability concepts in the maths curriculum. The bucket version is
simpler and more tactile---students cut up the input text (printed on
paper) using scissors and organise them into
buckets, making it suitable for younger learners or when
dice maths would be a distraction. Both variants teach the same core
concepts. The \url{www.llmsunplugged.org} website has a "toggle" for
these lessons to switch between the grid and bucket versions
of the activities.

\subsubsection{Extensions}

These lessons can be done in any order after completing the
Fundamentals. Each explores a different aspect of how modern language
models work. Most of them require a model that was created as part of
the \emph{Training} lesson, and students can use the one they created
earlier or swap models amongst themselves.

\paragraph{Scaling up}
\begin{itemize}
  \item
    \emph{Pre-trained Generation}: students use a provided booklet
    (generated via a website widget for any uploaded text)
    containing a model trained on a larger corpus, generating text
    without having trained the model themselves. This demonstrates the
    LLM-as-a-service model: most users never train their own models, they just
    use ones provided by others.
  \item
    \emph{Trigram}: students train a model tracking two-word
    contexts instead of one (this lesson has both grid and bucket
    variants). The generated text is ``better'', but this
    comes at a cost---significantly more grid rows or buckets to keep
    track of---illustrating the fundamental context-length tradeoff.
\end{itemize}

\paragraph{Controlling output}
\begin{itemize}
  \item
    \emph{Sampling}: students experiment with temperature (dividing
    counts before rolling) and truncation strategies (e.g. greedy, no-repeat,
    alliteration). The same model produces noticeably different outputs,
    showing that generation control matters as much as training data
    \citep{holtzmanCuriousCaseNeural2020}.
  \item
    \emph{Beam Search}: multiple students track parallel generation
    paths on separate papers, pruning to the top candidates after each
    step. This group activity shows why search strategies matter: beam
    search finds more coherent sequences than single-path sampling.
  \item
    \emph{Tool Use}: students designate people or objects as ``tools''
    with trigger words. When the model generates a trigger, generation
    pauses while the tool returns a result. This shows that LLMs don't
    contain all knowledge---they learn \emph{when} to delegate
    to external sources \citep{schickToolformerLanguageModels2023}.
\end{itemize}

\paragraph{Context and meaning}
\begin{itemize}
  \item
    \emph{Context Columns}: students add columns for grammatical
    categories (e.g. after verb, after pronoun, after preposition) and
    combine these counts with word-specific counts (the standard
    \emph{Generation} procedure) during generation.
    This hand-crafted attention mechanism previews how transformers
    learn to weight relevant context
    \citep{vaswaniAttentionAllYou2017}.
  \item
    \emph{Word Embeddings}: students treat each word's row as a vector
    and calculate distances between rows to build a similarity matrix.
    Words that behave alike cluster together, revealing that meaning
    emerges from patterns of usage
    \citep{mikolovEfficientEstimationWord2013,
    penningtonGloVeGlobalVectors2014}.
\end{itemize}

\paragraph{Model tuning}
\begin{itemize}
  \item
    \emph{LoRA}: students train a small ``adaptation grid'' (a
    smaller grid with only a subset of the rows) on new
    domain text and add its counts to their base model during
    generation. This shows how one foundation model can spawn thousands
    of specialised versions through lightweight add-on layers
    \citep{huLoRALowRankAdaptation2022}.
  \item
    \emph{RLHF}: students generate multiple candidate outputs, vote on
    preferences, then adjust counts ($+1$ for preferred transitions, $-1$
    for rejected ones). This shifts what ``good'' means from ``matches
    training data'' in the direction of ``matches human preferences''
    \citep{ouyangTrainingLanguageModels2022}.
  \item
    \emph{Synthetic Data}: students generate synthetic text from their
    model, train a new model on it, and compare. Patterns degrade across
    generations as rare words vanish and common phrases dominate,
    demonstrating why training on AI-generated content risks model collapse
    \citep{shumailovAICatchesGenerative2024}.
\end{itemize}

This progression covers key ideas used in frontier LLMs like ChatGPT,
Claude or Gemini,
albeit at smaller scale. The limited vocabulary and repetitive nature of
the children's books makes it feasible to explore all these concepts in
an unplugged fashion.

\subsection{Example lesson plan}
\label{sec:example-lesson-plan}

This 90-minute outline covers the two Fundamentals lessons
(\emph{Training} and
\emph{Generation}) plus the \emph{Pre-trained Generation} extension,
and has been
successfully delivered to groups from high-school age up and from 5
to 50 participants. The printed handouts associated with these three
lesson plans are included in an Appendix, although we encourage
educators to visit the website to see all the lesson materials and
instructor notes. If we have an additional 30 minutes (2 hours
total) then we often add the \emph{Sampling} lesson at the end.

\paragraph{Introduction (15 minutes)}
Icebreaker: ask participants to consider what it means to ``model
language'', followed by a show-of-hands poll revealing how recently
they've used ChatGPT.

\paragraph{Training (20 minutes)}
The educator demonstrates building a bigram model by reading a few
sentences from a children's book and filling in a grid with tally marks
showing which words follow which other words. Participants then work in
small groups (2--3 people) to train their own models on different pages
of text, experiencing firsthand how training means counting patterns.
The activity concludes by introducing key terminology (training, model,
token, vocabulary) and connecting the hands-on activity to the training
of ``real'' LLMs, giving a sense of the differences in scale between the
two processes.

\paragraph{Generation (20 minutes)}
Using a completed bigram grid the educator demonstrates text generation
by looking up possible next words, converting tally counts to dice
ranges, and rolling a die to make weighted random selections. Groups
then use their newly-trained models to generate sentences, producing
(sometimes) surprisingly coherent and/or delightfully nonsensical
outputs. Discussion introduces terminology (prompt, completion,
prediction) and emphasises that real LLMs generate text through the same
iterative sampling process.

\paragraph{Pre-trained Generation (20 minutes)}
Participants receive printed booklets containing bigram models
pre-trained on larger text corpora (typically 5000--10000 words). These
booklets allow immediate text generation without manual training. Groups
experiment with generating longer passages and compare outputs from
models trained on different genres (a fun variation is to not tell
participants what the training text was and have them guess). This
lesson introduces the concepts of pre-training and foundation models,
connecting to how LLMs are trained once on massive corpora then made
available for immediate use.

\paragraph{Closing (15 minutes)}
Summary and reflections: how has this workshop changed how you think
about language models? How has this changed how you will use language
models?

\section{Reception: notes from the field}

Over the past year at the Australian National University we have run
these \emph{LLMs Unplugged} activities with approximately 400
participants across many sessions in groups of five to fifty. The
participants have ranged from school-age to undergraduate
students (from all across campus) to senior executives in ``executive
education'' short courses. Interestingly, the majority of participants
have been senior leaders in the Australian Public Service---with a range
of different expertise and significant interest in understanding AI
tools they are being asked to use and evaluate.

This is a practitioner paper rather than a controlled pedagogical study.
We have not (yet) conducted pre/post testing of conceptual
understanding, run control groups without the intervention, or gathered
quantitative learning outcome data. What we can report is qualitative
reception across these diverse contexts.

The material is engaging and overwhelmingly well received. Participants
consistently report that the hands-on activity helps them to build a new
mental model of how LLMs work. The most common insight people articulate
is that LLMs are ``just'' doing probability and randomness at
scale---not reasoning, not understanding, but sophisticated pattern
matching and weighted sampling. This demystification seems particularly
valuable for non-technical participants who may have heard LLMs
described in almost magical terms.

The generative aspect matters. People are genuinely delighted when their
hand-built model produces a sentence that is both grammatical and
surprising. This is not just pedagogically useful---it is emotionally
engaging in a way that classification tasks are not. Several learners
have reported taking their models home to recreate the activity that
evening with their own teenage children. When your bigram model trained
on \emph{The Cat in the Hat} \citep{seuss1957cat} generates ``fish fish
fish red one fish two fish'', people laugh and immediately want to
generate more text to see what else might emerge.

One consistent observation about delivering this material is that
there is an inflection point after the first ``shareback'' of
the newly-generated text. The \emph{Training} lesson is necessary
set-up, but the room really starts to buzz during \emph{Generation}
when we go around the room
and people get to share what came out of their new model. Getting
each group to do a ``dramatic reading'' of their generated
text helps here too; the more they ham it up the better. From this
point on there are laughs and general
good vibes, and the questions they ask about LLMs are often more
incisive too. If
possible, we recommend facilitators leave enough space for an
engaging and playful ``shareback'' time when conducting the session.

The modular design has proven essential in practice. Most of our deliveries
are groups of five to fifty people and cover the Fundamentals lessons
plus \emph{Pre-trained
Generation}, with \emph{Sampling} added on to the end if we have two
hours instead of 90 minutes.
We have run all of the Extensions at least once, although
some are less ``battle-tested'' in different
classroom settings---something we are planning to rectify in the near future.
The ability to
scale the content up or down based on available time and audience
sophistication has been crucial for the material's adaptability.

% this is just here to ensure the acknowledgements stay on this page
% rather than getting pushed to the next one
\enlargethispage{1\baselineskip}
\section{Next steps}

The current material represents the beginning of an \emph{LLMs Unplugged}
resource pool, but there is plenty of room for more. As noted above,
most delivery focuses on the Fundamentals lessons (\emph{Training} and
\emph{Generation}) plus the \emph{Pre-trained Generation} and
\emph{Sampling} extensions.
These have been successfully run with groups from five to fifty people
in a tight 90-minute session (as per example lesson plan above),
covering the essential training-to-generation pipeline.

The Extension lessons go deeper, moving progressively closer
to how real
transformer models work while remaining unplugged. These lessons raise
interesting questions about depth in unplugged activities.
The CS Unplugged approach excels at providing intuitive introductions to
complex concepts---``aha'' moments that establish foundational
understanding \citep{bellCSUnpluggedHow2018}. But there is tension
between the constraints of unplugged activities (manual computation,
limited scale, short timeframe) and exploring concepts in depth. At some
point, does meaningful engagement with the ideas require moving beyond
the unplugged format?

Looking forward, we will continue to develop and deliver these lessons,
especially through teacher training. The entire project---lessons, printable
handouts, instructor notes, and software tools---will remain freely
available under a Creative Commons BY-NC-SA 4.0 license at
\url{www.llmsunplugged.org}. We encourage
educators to use, adapt, and improve the material, and we are interested
in hearing about implementations in different contexts.

Developing a mental model of how LLMs work should not require a computer
science degree or months of study. These unplugged activities
demonstrate that the core concepts are accessible to anyone willing to
spend an afternoon with pen, paper, and dice or scissors. As LLMs become
increasingly central to how we work with text and interact with digital
systems, this kind of hands-on understanding becomes not just
pedagogically valuable but practically necessary.

\emph{LLMs Unplugged} is not on its own going to save us from the impending
epistemological polycrisis \citep{strasserCarelessUseAI2025} or stop
people falling in love with chatbots \citep{liZhangFindingLove2024}.
Still, better and more widespread understanding of what LLMs are (and
aren't) is critical if we want people to make better informed decisions
about appropriate use and critical evaluation of their outputs.

\begin{acks}
  Many thanks to Eddie Aloise King and Cole Cooney for their input and
  help facilitating these activities over the past year.
\end{acks}

\bibliographystyle{ACM-Reference-Format}
\bibliography{ace-26.bib}

%%% -*-BibTeX-*-
%%% Do NOT edit. File created by BibTeX with style
%%% ACM-Reference-Format-Journals [18-Jan-2012].

\begin{thebibliography}{30}

%%% ====================================================================
%%% NOTE TO THE USER: you can override these defaults by providing
%%% customized versions of any of these macros before the \bibliography
%%% command.  Each of them MUST provide its own final punctuation,
%%% except for \shownote{} and \showURL{}.  The latter two
%%% do not use final punctuation, in order to avoid confusing it with
%%% the Web address.
%%%
%%% To suppress output of a particular field, define its macro to expand
%%% to an empty string, or better, \unskip, like this:
%%%
%%% \newcommand{\showURL}[1]{\unskip}   % LaTeX syntax
%%%
%%% \def \showURL #1{\unskip}           % plain TeX syntax
%%%
%%% ====================================================================

\ifx \showCODEN    \undefined \def \showCODEN     #1{\unskip}     \fi
\ifx \showISBNx    \undefined \def \showISBNx     #1{\unskip}     \fi
\ifx \showISBNxiii \undefined \def \showISBNxiii  #1{\unskip}     \fi
\ifx \showISSN     \undefined \def \showISSN      #1{\unskip}     \fi
\ifx \showLCCN     \undefined \def \showLCCN      #1{\unskip}     \fi
\ifx \shownote     \undefined \def \shownote      #1{#1}          \fi
\ifx \showarticletitle \undefined \def \showarticletitle #1{#1}   \fi
\ifx \showURL      \undefined \def \showURL       {\relax}        \fi
% The following commands are used for tagged output and should be
% invisible to TeX
\providecommand\bibfield[2]{#2}
\providecommand\bibinfo[2]{#2}
\providecommand\natexlab[1]{#1}
\providecommand\showeprint[2][]{arXiv:#2}

\bibitem[Bell and Vahrenhold(2018)]%
        {bellCSUnpluggedHow2018}
\bibfield{author}{\bibinfo{person}{Tim Bell} {and} \bibinfo{person}{Jan
  Vahrenhold}.} \bibinfo{year}{2018}\natexlab{}.
\newblock \showarticletitle{{{CS Unplugged}}---{{How Is It Used}}, and {{Does
  It Work}}?}
\newblock In \bibinfo{booktitle}{\emph{Adventures {{Between Lower Bounds}} and
  {{Higher Altitudes}}: {{Essays Dedicated}} to {{Juraj Hromkovi{\v c}}} on the
  {{Occasion}} of {{His}} 60th {{Birthday}}}},
  \bibfield{editor}{\bibinfo{person}{Hans-Joachim B{\"o}ckenhauer},
  \bibinfo{person}{Dennis Komm}, {and} \bibinfo{person}{Walter Unger}} (Eds.).
  \bibinfo{publisher}{Springer International Publishing},
  \bibinfo{address}{Cham}, \bibinfo{pages}{497--521}.
\newblock
\showISBNx{978-3-319-98355-4}
\href{https://doi.org/10.1007/978-3-319-98355-4_29}{doi:\nolinkurl{10.1007/978-3-319-98355-4_29}}


\bibitem[Campbell(1982)]%
        {campbell1982dear}
\bibfield{author}{\bibinfo{person}{Rod Campbell}.}
  \bibinfo{year}{1982}\natexlab{}.
\newblock \bibinfo{booktitle}{\emph{Dear Zoo: A Lift-the-Flap Book}}.
\newblock \bibinfo{publisher}{Blackie}, \bibinfo{address}{London}.
\newblock


\bibitem[Chen et~al\mbox{.}(2023)]%
        {chenFosteringComputationalThinking2023}
\bibfield{author}{\bibinfo{person}{Pei Chen}, \bibinfo{person}{Daner Yang},
  \bibinfo{person}{Ahmed Hosny~Saleh Metwally}, \bibinfo{person}{Jari Lavonen},
  {and} \bibinfo{person}{Xudong Wang}.} \bibinfo{year}{2023}\natexlab{}.
\newblock \showarticletitle{Fostering computational thinking through unplugged
  activities: A systematic literature review and meta-analysis}.
\newblock \bibinfo{journal}{\emph{International Journal of STEM Education}}
  \bibinfo{volume}{10}, \bibinfo{number}{1} (\bibinfo{year}{2023}),
  \bibinfo{pages}{47}.
\newblock
\href{https://doi.org/10.1186/s40594-023-00434-7}{doi:\nolinkurl{10.1186/s40594-023-00434-7}}


\bibitem[{Computer Science Education Research Group, University of
  Canterbury}({[n.\,d.]})]%
        {csunplugged}
\bibfield{author}{\bibinfo{person}{{Computer Science Education Research Group,
  University of Canterbury}}.} \bibinfo{year}{[n.\,d.]}\natexlab{}.
\newblock \bibinfo{title}{CS Unplugged}.
\newblock
\urldef\tempurl%
\url{https://www.csunplugged.org/en/}
\showURL{%
\tempurl}
\newblock
\shownote{Collection of freely available learning activities that teach
  computer science through engaging games and puzzles without using computers}.


\bibitem[Connelly et~al\mbox{.}(2025)]%
        {connellyLLMsBlackBoxes2025}
\bibfield{author}{\bibinfo{person}{Luke Connelly},
  \bibinfo{person}{Karl-Emil~Kj{\ae}r Bilstrup}, {and}
  \bibinfo{person}{Marianne~Graves Petersen}.} \bibinfo{year}{2025}\natexlab{}.
\newblock \showarticletitle{Beyond {{LLMs}} as {{Black Boxes}}: {{Activities}}
  and an {{Educational Tool Supporting Unplugged}} and {{Digital AI Learning
  Activities}} for {{K-12 Classrooms}}}. In \bibinfo{booktitle}{\emph{Adjunct
  {{Proceedings}} of the {{Sixth Decennial Aarhus Conference}}: {{Computing X
  Crisis}}}}. \bibinfo{publisher}{ACM}, \bibinfo{address}{Aarhus N Denmark},
  \bibinfo{pages}{1--4}.
\newblock
\showISBNx{979-8-4007-1968-4}
\href{https://doi.org/10.1145/3737609.3747112}{doi:\nolinkurl{10.1145/3737609.3747112}}


\bibitem[Gray and Sharp(1946)]%
        {gray1946fun}
\bibfield{author}{\bibinfo{person}{William~S. Gray} {and}
  \bibinfo{person}{Zerna Sharp}.} \bibinfo{year}{1946}\natexlab{}.
\newblock \bibinfo{booktitle}{\emph{Fun with Dick and Jane}}.
\newblock \bibinfo{publisher}{Scott, Foresman and Company},
  \bibinfo{address}{Chicago}.
\newblock


\bibitem[Holtzman et~al\mbox{.}(2020)]%
        {holtzmanCuriousCaseNeural2020}
\bibfield{author}{\bibinfo{person}{Ari Holtzman}, \bibinfo{person}{Jan Buys},
  \bibinfo{person}{Li Du}, \bibinfo{person}{Maxwell Forbes}, {and}
  \bibinfo{person}{Yejin Choi}.} \bibinfo{year}{2020}\natexlab{}.
\newblock \showarticletitle{The Curious Case of Neural Text Degeneration}. In
  \bibinfo{booktitle}{\emph{8th International Conference on Learning
  Representations, {ICLR} 2020, Addis Ababa, Ethiopia, April 26-30, 2020}}.
  \bibinfo{publisher}{OpenReview.net}.
\newblock
\urldef\tempurl%
\url{https://openreview.net/forum?id=rygGQyrFvH}
\showURL{%
\tempurl}


\bibitem[Hu et~al\mbox{.}(2022)]%
        {huLoRALowRankAdaptation2022}
\bibfield{author}{\bibinfo{person}{Edward~J. Hu}, \bibinfo{person}{Yelong
  Shen}, \bibinfo{person}{Phillip Wallis}, \bibinfo{person}{Zeyuan
  {Allen-Zhu}}, \bibinfo{person}{Yuanzhi Li}, \bibinfo{person}{Shean Wang},
  \bibinfo{person}{Lu Wang}, {and} \bibinfo{person}{Weizhu Chen}.}
  \bibinfo{year}{2022}\natexlab{}.
\newblock \showarticletitle{{{LoRA}}: {{Low-Rank Adaptation}} of {{Large
  Language Models}}}. In \bibinfo{booktitle}{\emph{The Tenth International
  Conference on Learning Representations, {ICLR} 2022, Virtual Event, April
  25-29, 2022}}. \bibinfo{publisher}{OpenReview.net}.
\newblock
\urldef\tempurl%
\url{https://openreview.net/forum?id=nZeVKeeFYf9}
\showURL{%
\tempurl}


\bibitem[Li and Zhang(2024)]%
        {liZhangFindingLove2024}
\bibfield{author}{\bibinfo{person}{Han Li} {and} \bibinfo{person}{Renwen
  Zhang}.} \bibinfo{year}{2024}\natexlab{}.
\newblock \showarticletitle{Finding Love in Algorithms: Deciphering the
  Emotional Contexts of Close Encounters with {{AI}} Chatbots}.
\newblock \bibinfo{journal}{\emph{Journal of Computer-Mediated Communication}}
  \bibinfo{volume}{29}, \bibinfo{number}{5} (\bibinfo{date}{Aug.}
  \bibinfo{year}{2024}), \bibinfo{pages}{zmae015}.
\newblock
\href{https://doi.org/10.1093/jcmc/zmae015}{doi:\nolinkurl{10.1093/jcmc/zmae015}}


\bibitem[Liao et~al\mbox{.}(2025)]%
        {liaoLLMsResearchTools2025}
\bibfield{author}{\bibinfo{person}{Zhehui Liao}, \bibinfo{person}{Maria
  Antoniak}, \bibinfo{person}{Inyoung Cheong}, \bibinfo{person}{Evie Yu-Yen
  Cheng}, \bibinfo{person}{Ai-Heng Lee}, \bibinfo{person}{Kyle Lo},
  \bibinfo{person}{Joseph~Chee Chang}, {and} \bibinfo{person}{Amy~X. Zhang}.}
  \bibinfo{year}{2025}\natexlab{}.
\newblock \showarticletitle{{{LLMs}} as {{Research Tools}}: {{A Large Scale
  Survey}} of {{Researchers}}' {{Usage}} and {{Perceptions}}}. In
  \bibinfo{booktitle}{\emph{Second Conference on Language Modeling, {COLM}
  2025, Montreal, Canada, October 7-10, 2025}}.
  \bibinfo{publisher}{OpenReview.net}.
\newblock
\urldef\tempurl%
\url{https://openreview.net/forum?id=p0BwJk3R1p}
\showURL{%
\tempurl}


\bibitem[Lindner et~al\mbox{.}(2019)]%
        {lindnerUnpluggedActivitiesContext2019}
\bibfield{author}{\bibinfo{person}{Annabel Lindner}, \bibinfo{person}{Stefan
  Seegerer}, {and} \bibinfo{person}{Ralf Romeike}.}
  \bibinfo{year}{2019}\natexlab{}.
\newblock \showarticletitle{Unplugged {{Activities}} in the {{Context}} of
  {{AI}}}. In \bibinfo{booktitle}{\emph{Informatics in {{Schools}}. {{New
  Ideas}} in {{School Informatics}}}},
  \bibfield{editor}{\bibinfo{person}{Sergei~N. Pozdniakov} {and}
  \bibinfo{person}{Valentina Dagien{\.e}}} (Eds.). \bibinfo{publisher}{Springer
  International Publishing}, \bibinfo{address}{Cham},
  \bibinfo{pages}{123--135}.
\newblock
\showISBNx{978-3-030-33759-9}
\href{https://doi.org/10.1007/978-3-030-33759-9_10}{doi:\nolinkurl{10.1007/978-3-030-33759-9_10}}


\bibitem[Link(2006)]%
        {linkChainsWestMarkovs2006}
\bibfield{author}{\bibinfo{person}{David Link}.}
  \bibinfo{year}{2006}\natexlab{}.
\newblock \showarticletitle{Chains to the {{West}}: {{Markov}}'s {{Theory}} of
  {{Connected Events}} and {{Its Transmission}} to {{Western Europe}}}.
\newblock \bibinfo{journal}{\emph{Science in Context}} \bibinfo{volume}{19},
  \bibinfo{number}{4} (\bibinfo{date}{Dec.} \bibinfo{year}{2006}),
  \bibinfo{pages}{561--589}.
\newblock
\showISSN{0269-8897, 1474-0664}
\href{https://doi.org/10.1017/S0269889706001062}{doi:\nolinkurl{10.1017/S0269889706001062}}


\bibitem[Matsakis and Klock(2014)]%
        {matsakisRustLanguage2014}
\bibfield{author}{\bibinfo{person}{Nicholas~D. Matsakis} {and}
  \bibinfo{person}{II Klock, Felix~S.}} \bibinfo{year}{2014}\natexlab{}.
\newblock \showarticletitle{The Rust Language}. In
  \bibinfo{booktitle}{\emph{Proceedings of the {{ACM SIGAda Annual Conference}}
  on {{High Integrity Language Technology}}}}. \bibinfo{publisher}{ACM},
  \bibinfo{address}{Portland, OR, USA}, \bibinfo{pages}{103--104}.
\newblock
\showISBNx{978-1-4503-3217-0}
\href{https://doi.org/10.1145/2663171.2663188}{doi:\nolinkurl{10.1145/2663171.2663188}}


\bibitem[Mikolov et~al\mbox{.}(2013)]%
        {mikolovEfficientEstimationWord2013}
\bibfield{author}{\bibinfo{person}{Tomas Mikolov}, \bibinfo{person}{Kai Chen},
  \bibinfo{person}{Greg Corrado}, {and} \bibinfo{person}{Jeffrey Dean}.}
  \bibinfo{year}{2013}\natexlab{}.
\newblock \showarticletitle{Efficient Estimation of Word Representations in
  Vector Space}. In \bibinfo{booktitle}{\emph{1st International Conference on
  Learning Representations, {ICLR} 2013, Scottsdale, Arizona, USA, May 2-4,
  2013, Workshop Track Proceedings}}.
\newblock
\urldef\tempurl%
\url{http://arxiv.org/abs/1301.3781}
\showURL{%
\tempurl}


\bibitem[Mädje(2022)]%
        {maedjeTypstProgrammableMarkup2022}
\bibfield{author}{\bibinfo{person}{Laurenz Mädje}.}
  \bibinfo{year}{2022}\natexlab{}.
\newblock \emph{\bibinfo{title}{A Programmable Markup Language for
  Typesetting}}.
\newblock \bibinfo{thesistype}{Master's\ thesis}. \bibinfo{school}{Technische
  Universität Berlin}, \bibinfo{address}{Berlin, Germany}.
\newblock
\urldef\tempurl%
\url{https://laurmaedje.github.io/programmable-markup-language-for-typesetting.pdf}
\showURL{%
\tempurl}


\bibitem[Nishida et~al\mbox{.}(2009)]%
        {nishidaCSUnpluggedDesign2009}
\bibfield{author}{\bibinfo{person}{Tomohiro Nishida}, \bibinfo{person}{Susumu
  Kanemune}, \bibinfo{person}{Yukio Idosaka}, \bibinfo{person}{Mitaro Namiki},
  \bibinfo{person}{Timothy~C. Bell}, {and} \bibinfo{person}{Yasushi Kuno}.}
  \bibinfo{year}{2009}\natexlab{}.
\newblock \showarticletitle{A {{CS}} Unplugged Design Pattern}. In
  \bibinfo{booktitle}{\emph{Proceedings of the 40th {{ACM Technical Symposium}}
  on {{Computer Science Education}}}}. \bibinfo{publisher}{ACM},
  \bibinfo{address}{Chattanooga, TN, USA}, \bibinfo{pages}{231--235}.
\newblock
\href{https://doi.org/10.1145/1508865.1508951}{doi:\nolinkurl{10.1145/1508865.1508951}}


\bibitem[{Northwestern University}({[n.\,d.]})]%
        {northwesternAIUnplugged}
\bibfield{author}{\bibinfo{person}{{Northwestern University}}.}
  \bibinfo{year}{[n.\,d.]}\natexlab{}.
\newblock \bibinfo{title}{AI Unplugged}.
\newblock
\urldef\tempurl%
\url{https://sites.northwestern.edu/aiunplugged/}
\showURL{%
\tempurl}
\newblock
\shownote{Collection of no-computer AI literacy activities for middle-school
  students, including lesson plans on machine learning, data privacy, and AI
  ethics. Funded by NSF DRL 2343693}.


\bibitem[{OpenAI}(2022)]%
        {openaiChatGPT2022}
\bibfield{author}{\bibinfo{person}{{OpenAI}}.} \bibinfo{year}{2022}\natexlab{}.
\newblock \bibinfo{title}{ChatGPT}.
\newblock
\urldef\tempurl%
\url{https://openai.com/index/chatgpt/}
\showURL{%
\tempurl}
\newblock
\shownote{Conversational AI system launched November 30, 2022}.


\bibitem[Ouyang et~al\mbox{.}(2022)]%
        {ouyangTrainingLanguageModels2022}
\bibfield{author}{\bibinfo{person}{Long Ouyang}, \bibinfo{person}{Jeffrey Wu},
  \bibinfo{person}{Xu Jiang}, \bibinfo{person}{Diogo Almeida},
  \bibinfo{person}{Carroll Wainwright}, \bibinfo{person}{Pamela Mishkin},
  \bibinfo{person}{Chong Zhang}, \bibinfo{person}{Sandhini Agarwal},
  \bibinfo{person}{Katarina Slama}, \bibinfo{person}{Alex Ray},
  \bibinfo{person}{John Schulman}, \bibinfo{person}{Jacob Hilton},
  \bibinfo{person}{Fraser Kelton}, \bibinfo{person}{Luke Miller},
  \bibinfo{person}{Maddie Simens}, \bibinfo{person}{Amanda Askell},
  \bibinfo{person}{Peter Welinder}, \bibinfo{person}{Paul~F. Christiano},
  \bibinfo{person}{Jan Leike}, {and} \bibinfo{person}{Ryan Lowe}.}
  \bibinfo{year}{2022}\natexlab{}.
\newblock \showarticletitle{Training Language Models to Follow Instructions
  with Human Feedback}. In \bibinfo{booktitle}{\emph{Advances in Neural
  Information Processing Systems}}, Vol.~\bibinfo{volume}{35}.
  \bibinfo{publisher}{Curran Associates, Inc.}, \bibinfo{pages}{27730--27744}.
\newblock
\urldef\tempurl%
\url{https://proceedings.neurips.cc/paper_files/paper/2022/hash/b1efde53be364a73914f58805a001731-Abstract-Conference.html}
\showURL{%
\tempurl}


\bibitem[Pennington et~al\mbox{.}(2014)]%
        {penningtonGloVeGlobalVectors2014}
\bibfield{author}{\bibinfo{person}{Jeffrey Pennington},
  \bibinfo{person}{Richard Socher}, {and} \bibinfo{person}{Christopher
  Manning}.} \bibinfo{year}{2014}\natexlab{}.
\newblock \showarticletitle{{{GloVe}}: {{Global Vectors}} for {{Word
  Representation}}}. In \bibinfo{booktitle}{\emph{Proceedings of the 2014
  Conference on Empirical Methods in Natural Language Processing ({{EMNLP}})}}.
  \bibinfo{publisher}{Association for Computational Linguistics},
  \bibinfo{address}{Doha, Qatar}, \bibinfo{pages}{1532--1543}.
\newblock
\href{https://doi.org/10.3115/v1/D14-1162}{doi:\nolinkurl{10.3115/v1/D14-1162}}


\bibitem[Schick et~al\mbox{.}(2023)]%
        {schickToolformerLanguageModels2023}
\bibfield{author}{\bibinfo{person}{Timo Schick}, \bibinfo{person}{Jane
  Dwivedi-Yu}, \bibinfo{person}{Roberto Dess{\`i}}, \bibinfo{person}{Roberta
  Raileanu}, \bibinfo{person}{Maria Lomeli}, \bibinfo{person}{Luke
  Zettlemoyer}, \bibinfo{person}{Nicola Cancedda}, {and}
  \bibinfo{person}{Thomas Scialom}.} \bibinfo{year}{2023}\natexlab{}.
\newblock \showarticletitle{Toolformer: {{Language Models Can Teach
  Themselves}} to {{Use Tools}}}. In \bibinfo{booktitle}{\emph{Advances in
  Neural Information Processing Systems}}, Vol.~\bibinfo{volume}{36}.
  \bibinfo{publisher}{Curran Associates, Inc.}, \bibinfo{pages}{68539--68551}.
\newblock
\urldef\tempurl%
\url{https://proceedings.neurips.cc/paper_files/paper/2023/hash/d842425e4bf79ba039352da0f658a906-Abstract-Conference.html}
\showURL{%
\tempurl}


\bibitem[Seegerer and Lindner({[n.\,d.]})]%
        {seegererAIUnplugged}
\bibfield{author}{\bibinfo{person}{Stefan Seegerer} {and}
  \bibinfo{person}{Annabel Lindner}.} \bibinfo{year}{[n.\,d.]}\natexlab{}.
\newblock \bibinfo{title}{AI Unplugged}.
\newblock
\urldef\tempurl%
\url{https://www.aiunplugged.org}
\showURL{%
\tempurl}
\newblock
\shownote{Activities and teaching material on artificial intelligence,
  including downloadable board games and activities in multiple languages}.


\bibitem[Seuss(1957)]%
        {seuss1957cat}
\bibfield{author}{\bibinfo{person}{Dr. Seuss}.}
  \bibinfo{year}{1957}\natexlab{}.
\newblock \bibinfo{booktitle}{\emph{The Cat in the Hat}}.
\newblock \bibinfo{publisher}{Random House}, \bibinfo{address}{New York}.
\newblock


\bibitem[Shannon(1951)]%
        {shannonPredictionEntropyPrinted1951}
\bibfield{author}{\bibinfo{person}{Claude~E. Shannon}.}
  \bibinfo{year}{1951}\natexlab{}.
\newblock \showarticletitle{Prediction and {{Entropy}} of {{Printed English}}}.
\newblock \bibinfo{journal}{\emph{Bell System Technical Journal}}
  \bibinfo{volume}{30}, \bibinfo{number}{1} (\bibinfo{date}{Jan.}
  \bibinfo{year}{1951}), \bibinfo{pages}{50--64}.
\newblock
\href{https://doi.org/10.1002/j.1538-7305.1951.tb01366.x}{doi:\nolinkurl{10.1002/j.1538-7305.1951.tb01366.x}}


\bibitem[Shumailov et~al\mbox{.}(2024)]%
        {shumailovAICatchesGenerative2024}
\bibfield{author}{\bibinfo{person}{Ilia Shumailov}, \bibinfo{person}{Zakhar
  Shumaylov}, \bibinfo{person}{Yiren Zhao}, \bibinfo{person}{Yarin Gal},
  \bibinfo{person}{Nicolas Papernot}, {and} \bibinfo{person}{Ross Anderson}.}
  \bibinfo{year}{2024}\natexlab{}.
\newblock \showarticletitle{{{AI}} Models Collapse When Trained on Recursively
  Generated Data}.
\newblock \bibinfo{journal}{\emph{Nature}} \bibinfo{volume}{631},
  \bibinfo{number}{8022} (\bibinfo{date}{July} \bibinfo{year}{2024}),
  \bibinfo{pages}{755--759}.
\newblock
\showISSN{1476-4687}
\href{https://doi.org/10.1038/s41586-024-07566-y}{doi:\nolinkurl{10.1038/s41586-024-07566-y}}


\bibitem[Song et~al\mbox{.}(2024)]%
        {songArtificialIntelligenceUnplugged2024}
\bibfield{author}{\bibinfo{person}{Yukyeong Song}, \bibinfo{person}{Xiaoyi
  Tian}, \bibinfo{person}{Nandika Regatti}, \bibinfo{person}{Gloria~Ashiya
  Katuka}, \bibinfo{person}{Kristy~Elizabeth Boyer}, {and}
  \bibinfo{person}{Maya Israel}.} \bibinfo{year}{2024}\natexlab{}.
\newblock \showarticletitle{Artificial {{Intelligence Unplugged}}: {{Designing
  Unplugged Activities}} for a {{Conversational AI Summer Camp}}}. In
  \bibinfo{booktitle}{\emph{Proceedings of the 55th {{ACM Technical Symposium}}
  on {{Computer Science Education V}}. 1}}. \bibinfo{publisher}{ACM},
  \bibinfo{address}{Portland OR USA}, \bibinfo{pages}{1272--1278}.
\newblock
\showISBNx{979-8-4007-0423-9}
\href{https://doi.org/10.1145/3626252.3630783}{doi:\nolinkurl{10.1145/3626252.3630783}}


\bibitem[Strasser(2025)]%
        {strasserCarelessUseAI2025}
\bibfield{author}{\bibinfo{person}{Anna Strasser}.}
  \bibinfo{year}{2025}\natexlab{}.
\newblock \showarticletitle{Why a Careless Use of {{AI}} Tools May Contribute
  to an Epistemological Crisis}.
\newblock \bibinfo{journal}{\emph{P\&D - Philosophy \& Digitality}}
  \bibinfo{volume}{2}, \bibinfo{number}{1} (\bibinfo{year}{2025}),
  \bibinfo{pages}{162--184}.
\newblock
\href{https://doi.org/10.18716/pd.v2i1.11662}{doi:\nolinkurl{10.18716/pd.v2i1.11662}}


\bibitem[Vaswani et~al\mbox{.}(2017)]%
        {vaswaniAttentionAllYou2017}
\bibfield{author}{\bibinfo{person}{Ashish Vaswani}, \bibinfo{person}{Noam
  Shazeer}, \bibinfo{person}{Niki Parmar}, \bibinfo{person}{Jakob Uszkoreit},
  \bibinfo{person}{Llion Jones}, \bibinfo{person}{Aidan~N. Gomez},
  \bibinfo{person}{Lukasz Kaiser}, {and} \bibinfo{person}{Illia Polosukhin}.}
  \bibinfo{year}{2017}\natexlab{}.
\newblock \showarticletitle{Attention Is All You Need}. In
  \bibinfo{booktitle}{\emph{Advances in Neural Information Processing
  Systems}}, Vol.~\bibinfo{volume}{30}. \bibinfo{publisher}{Curran Associates,
  Inc.}, \bibinfo{address}{Red Hook, NY, USA}, \bibinfo{pages}{5998--6008}.
\newblock
\urldef\tempurl%
\url{https://proceedings.neurips.cc/paper_files/paper/2017/file/3f5ee243547dee91fbd053c1c4a845aa-Paper.pdf}
\showURL{%
\tempurl}


\bibitem[Williams et~al\mbox{.}(2020)]%
        {williamsCSInSchools2020}
\bibfield{author}{\bibinfo{person}{Hugh~E. Williams}, \bibinfo{person}{Selina
  Williams}, {and} \bibinfo{person}{Kristy Kendall}.}
  \bibinfo{year}{2020}\natexlab{}.
\newblock \showarticletitle{{{CS}} in {{Schools}}: {{Developing}} a
  {{Sustainable Coding Programme}} in {{Australian Schools}}}. In
  \bibinfo{booktitle}{\emph{Proceedings of the 2020 {{ACM Conference}} on
  {{Innovation}} and {{Technology}} in {{Computer Science Education}}}}.
  \bibinfo{publisher}{ACM}, \bibinfo{address}{Trondheim, Norway},
  \bibinfo{pages}{321--327}.
\newblock
\href{https://doi.org/10.1145/3341525.3387422}{doi:\nolinkurl{10.1145/3341525.3387422}}


\bibitem[Zhang and Tur(2024)]%
        {zhangSystematicReviewChatGPT2024}
\bibfield{author}{\bibinfo{person}{Peng Zhang} {and} \bibinfo{person}{Gemma
  Tur}.} \bibinfo{year}{2024}\natexlab{}.
\newblock \showarticletitle{A Systematic Review of {{{\textsc{ChatGPT}}}} Use
  in {{K}}-12 Education}.
\newblock \bibinfo{journal}{\emph{European Journal of Education}}
  \bibinfo{volume}{59}, \bibinfo{number}{2} (\bibinfo{date}{June}
  \bibinfo{year}{2024}), \bibinfo{pages}{e12599}.
\newblock
\showISSN{0141-8211, 1465-3435}
\href{https://doi.org/10.1111/ejed.12599}{doi:\nolinkurl{10.1111/ejed.12599}}


\end{thebibliography}

\appendix
\section*{Appendix: lesson cards}

The following pages show printable lesson handouts for the
lesson plan described in Section~\ref{sec:example-lesson-plan}. Each
card is designed to be printed on double-sided A4 paper and includes
the lesson goal, required materials, step-by-step
procedure, and a worked example. The full set of handouts (as well as
  instructor notes, interactive demo widgets, and tools for making custom
"pre-trained language model" booklets) can be
found on the \url{www.llmsunplugged.org} website.

\onecolumn
%\newgeometry{margin=1cm}

% \pagestyle{empty}
\aptLtoX[graphic=no,type=html]{\centering
  \includegraphics[width=0.80\linewidth]{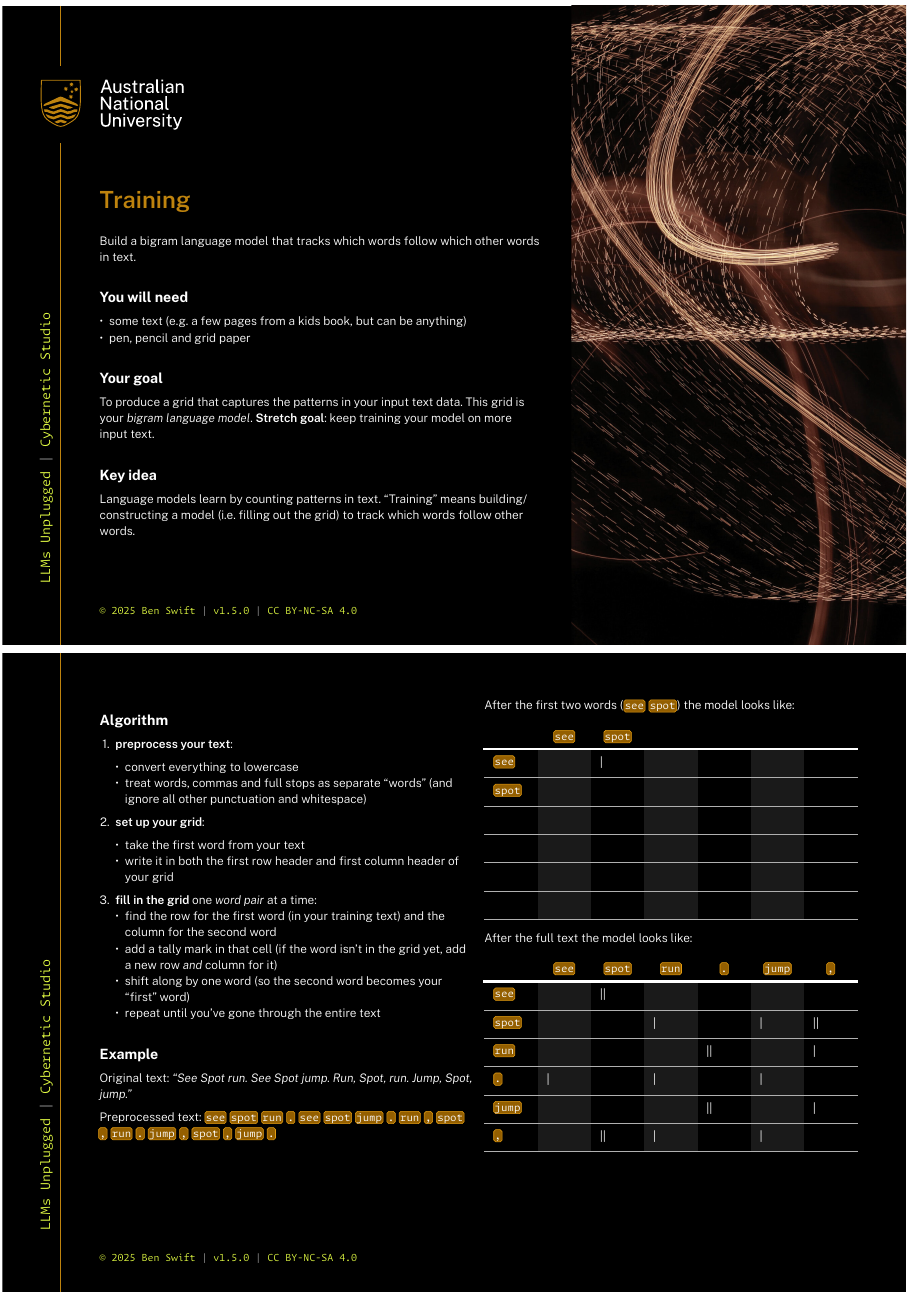}

  \includegraphics[width=0.80\linewidth]{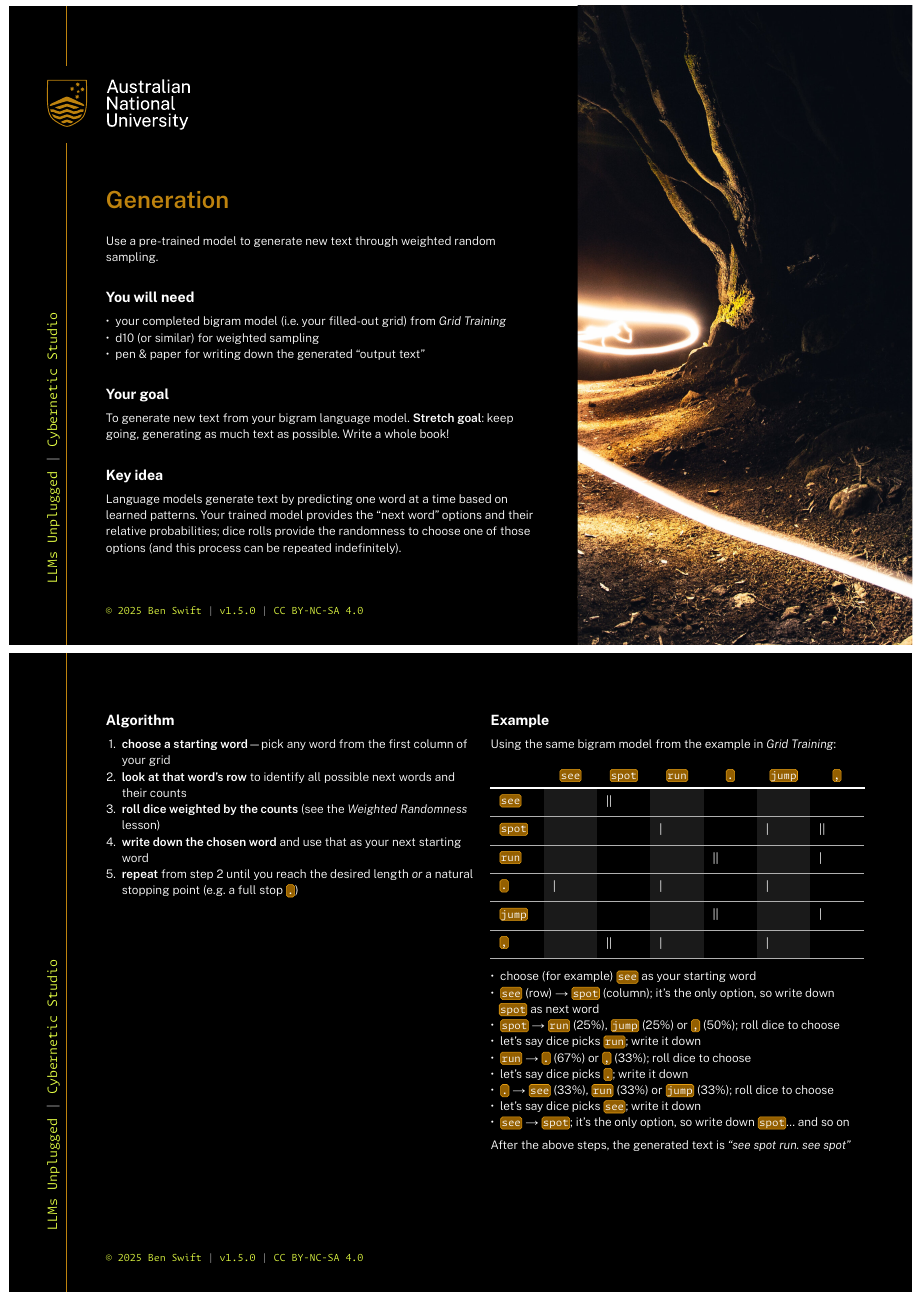}

\includegraphics[width=0.80\linewidth]{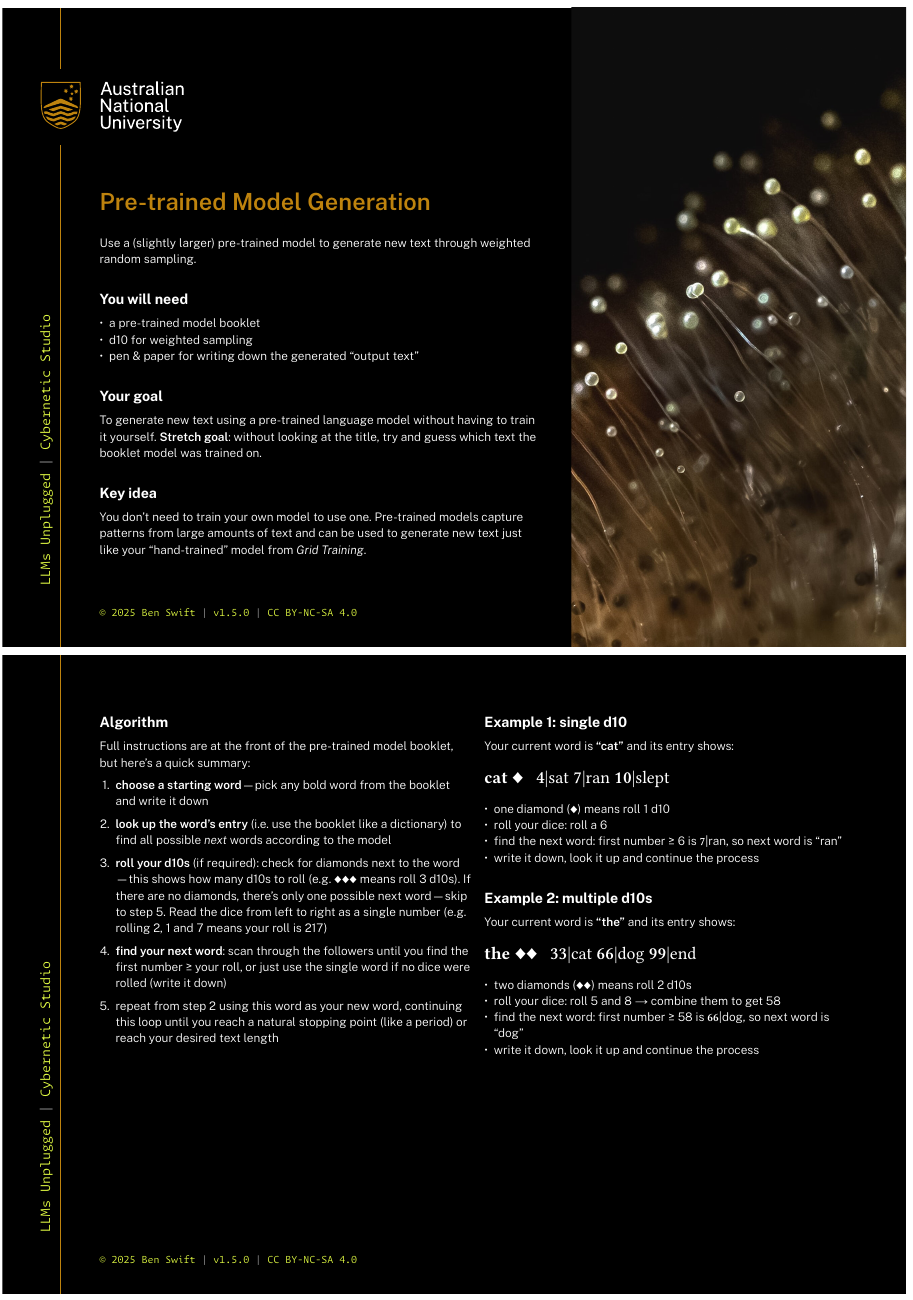}}{
  \centering
  \includegraphics[page=1, width=0.86\linewidth, alt={Training lesson
      card page 1: Instructions for building a bigram language model by
      counting word pairs. Shows the goal (produce a grid capturing
      patterns in input text), required materials (text, pen, grid paper),
      and the key idea that language models learn by counting which words
  follow other words.}]{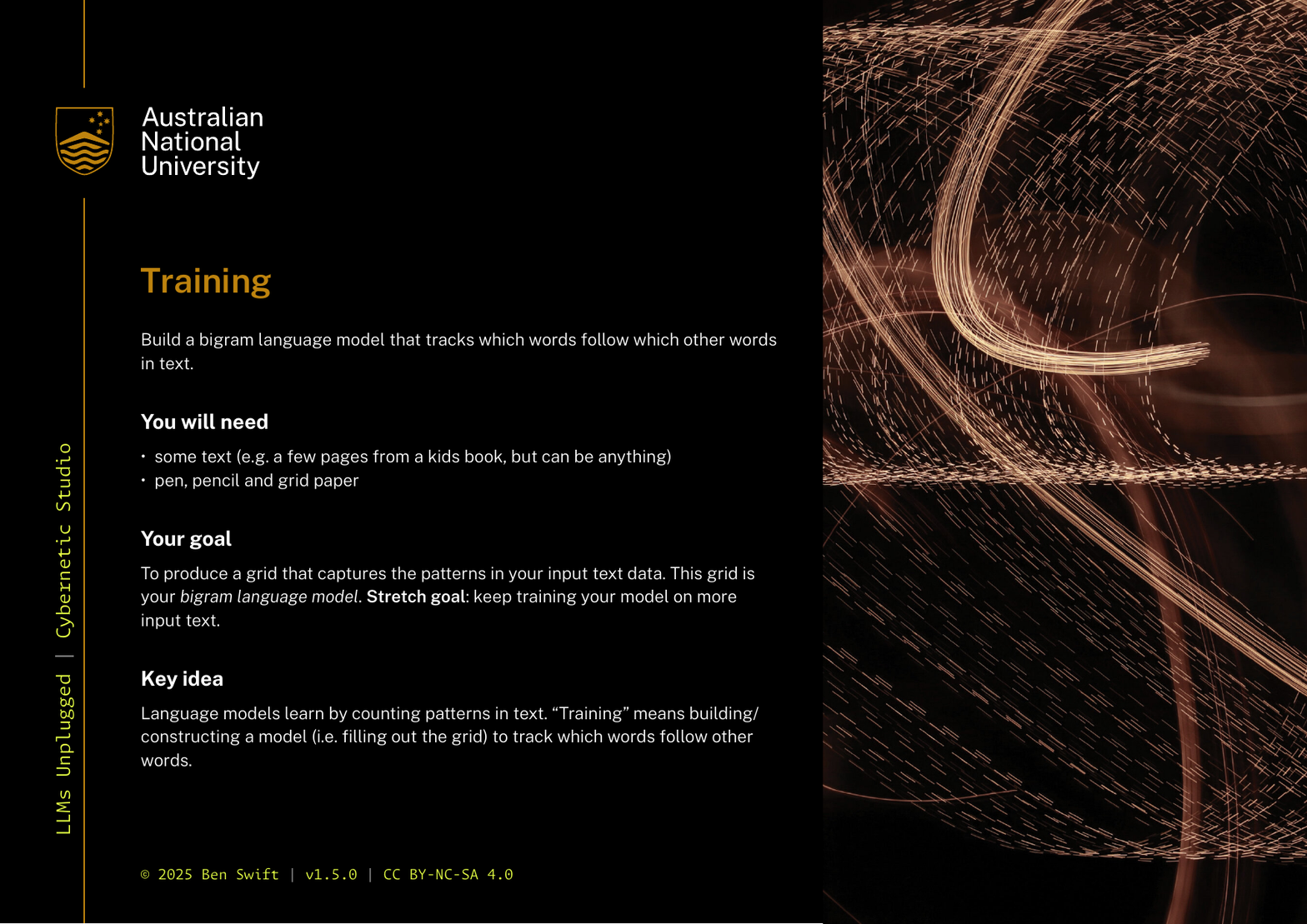}

  \vspace{0.1cm}
  \includegraphics[page=2, width=0.86\linewidth, alt={Training lesson
      card page 2: Step-by-step algorithm for filling the bigram grid.
      Includes preprocessing steps (lowercase, separate punctuation), grid
      setup, and the counting procedure. Shows a worked example using the
      text "See Spot run. See Spot jump." with the resulting bigram
  frequency grid.}]{grid-training.pdf}
  %\restoregeometry

  %\newgeometry{margin=1cm}
  \centering
  \includegraphics[page=1, width=0.86\linewidth, alt={Generation lesson
      card page 1: Instructions for generating text from a trained bigram
      model using weighted random sampling. Shows the goal (generate new
      text from the model), required materials (completed grid, d10 dice,
      pen and paper), and the key idea that models predict one word at a
  time based on learned patterns.}]{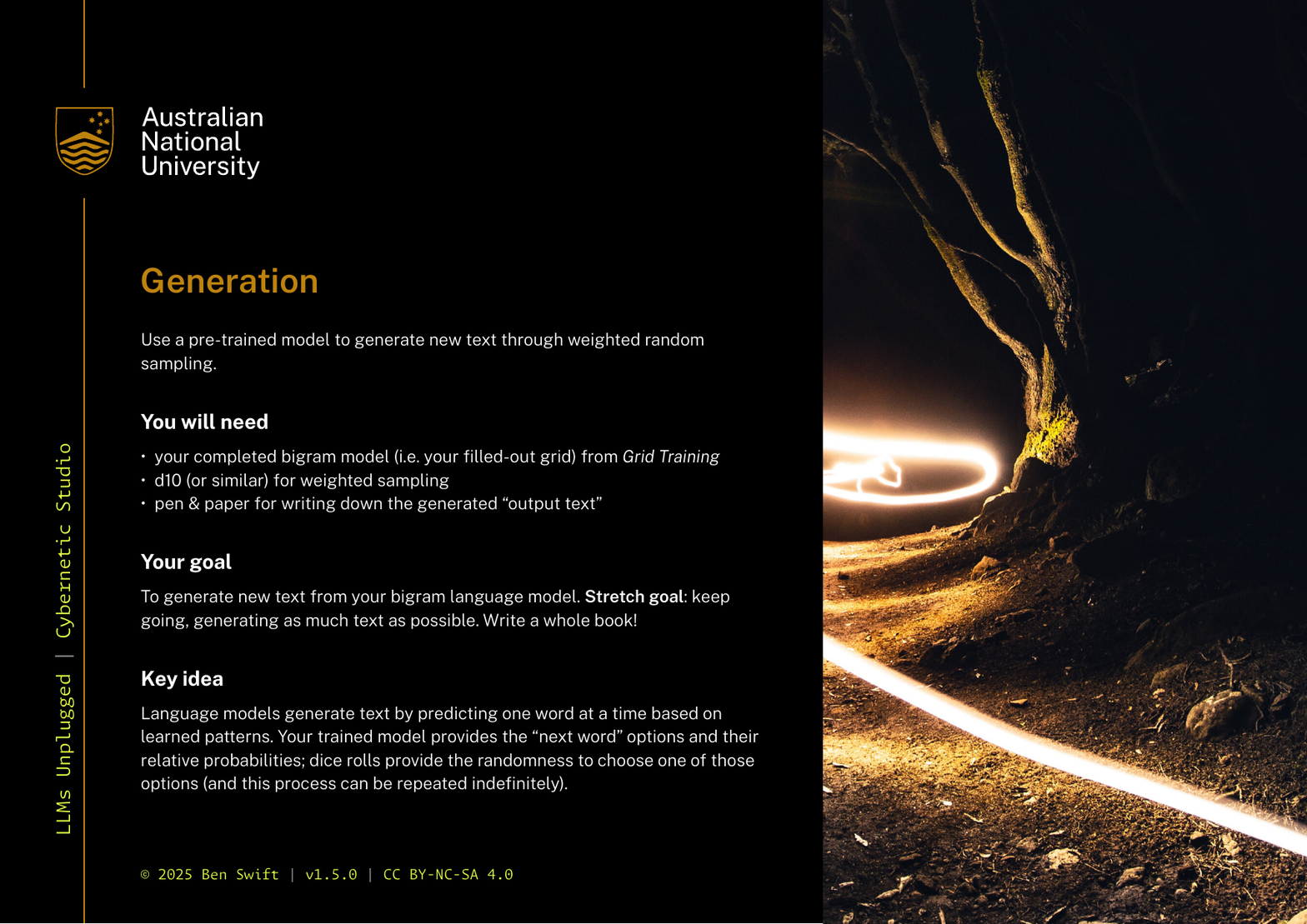}

  \vspace{0.1cm}
  \includegraphics[page=2, width=0.86\linewidth, alt={Generation lesson
      card page 2: Step-by-step algorithm for text generation. Choose a
      starting word, look up possible next words and their counts, roll
      dice weighted by counts, write down the chosen word, and repeat.
      Includes a worked example generating "see spot run. see spot" from
  the bigram model.}]{grid-generation.pdf}
  %\restoregeometry

  %\newgeometry{margin=1cm}
  \centering
  \includegraphics[page=1, width=0.86\linewidth, alt={Pre-trained
      generation lesson card page 1: Instructions for using a pre-trained
      model booklet to generate text. Shows the goal (generate text without
      training your own model), required materials (pre-trained booklet,
      d10 dice, pen and paper), and the key idea that pre-trained models
  capture patterns from large text corpora.}]{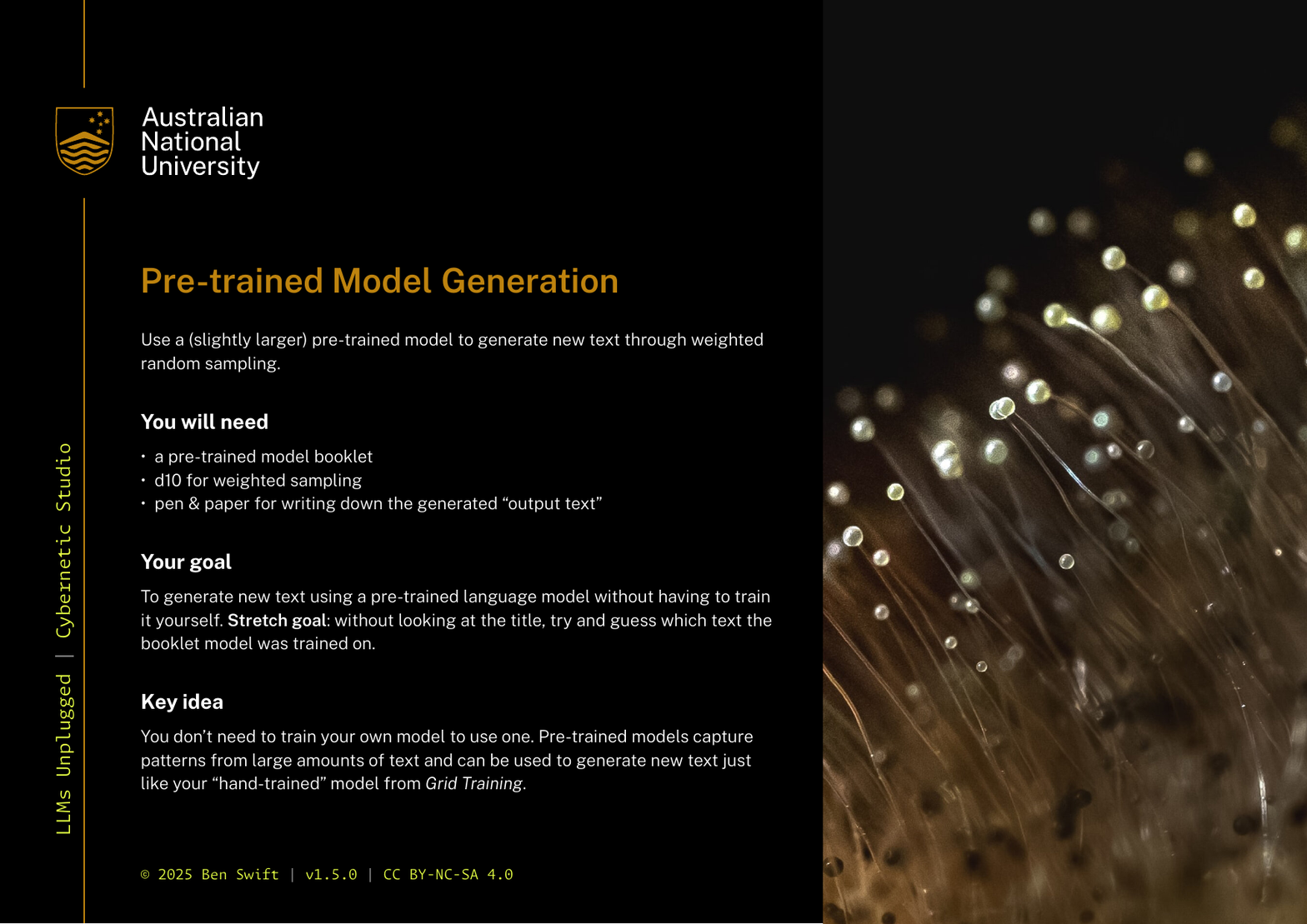}

  \vspace{0.1cm}
  \includegraphics[page=2, width=0.86\linewidth, alt={Pre-trained
      generation lesson card page 2: Algorithm for using the booklet
      format. Look up words like a dictionary, count diamonds to determine
      dice rolls, roll d10s and read as a number, find the first entry
      greater than or equal to your roll. Includes examples for single and
multiple d10 rolls.}]{pretrained-generation.pdf}}
%\restoregeometry

\end{document}